\documentclass[
    ,final            
  ]
  {aipproc}

\layoutstyle{8x11double}

\begin{document}

\title{Centrifugally induced curvature drift instability in AGN}


\author{Osmanov Zaza}{
  address={E. Kharadze Georgian National Astrophysical Observatory,
              Kazbegi str. 2a, 0106 Tbilisi, Georgia\\
              {email: z.osmanov@astro-ge.org}}
}



\begin{abstract}
We investigate the centrifugally driven curvature drift instability
   to study how field lines twist close to the light
   cylinder surface of an AGN, through which the free motion of AGN
   winds can be monitored.
By studying the dynamics of the relativistic MHD flow close to the
light
    cylinder surface, we derive and solve analytically the dispersion
    relation of the instability by applying a single particle approach
    based on the centrifugal acceleration.
Considering the typical values
   of AGN winds, it is shown that the timescale of the curvature drift instability
   is far less than the accretion process timescale, indicating that
   the present instability is very efficient and might strongly
   influence processes in AGN plasmas.
\end{abstract}

\maketitle


\section{Introduction}

For studying AGN winds the fundamental problem relates to the
understanding of a question: how the plasma goes through the Light
Cylinder Surface (LCS), which is the hypothetical zone where the
linear velocity of rotation equals the speed of light. This implies
that the plasma particles, which move along quasi-straight magnetic
field lines in the nearby area of the LCS, must reach the speed of
light. Generally speaking no physical system can maintain such a
motion and a certain twisting process of the magnetic field lines
must operate on the LCS. On the other hand if the trajectories are
given by the Archimedes spiral, then the particles can cross the LCS
avoiding the light cylinder problem \cite{r03}. An additional step
in this investigation is to identify the appropriate mechanism that
provides the twisting of the magnetic field lines, giving rise to
the shape of the Archimedes spiral, and in turn insures that the
dynamics is force-free.

Since the innermost region of AGNs rotates, the role of the
Centrifugal Force (CF) appears interesting. The centrifugally driven
outflows have been extensively studied. Generalizing the work
developed in \cite{gl97} it was shown that due to the centrifugal
acceleration, electrons gain very high energies with Lorentz factors
up to $\gamma\sim 10^8$ \cite{osm7,ra8}. This implies that the
energy budget in the AGN winds is very high.

The centrifugally driven parametric instability was first introduced
in \cite{incr1,incr3} for the Crab pulsar and AGN jets respectively.
Another kind of the instability which might be induced by the CF is
the so called Curvature Drift Instability (CDI). In \cite{mnras} the
two-component relativistic plasma has been considered to study the
role of the centrifugal acceleration in the curvature drift
instability for pulsar magnetospheres. The investigation
demonstrated high efficiency of the CDI. To investigate the twisting
process of magnetic field lines due to the CDI, we apply the method
developed in \cite{mnras,ff} to AGN winds.

The paper is arranged as follows. In Sect. 2, we introduce the
curvature drift waves and derive the dispersion relation. In Sect.
3, the results for typical AGNs are presented and, in Sect. 4, we
summarize our results.
\begin{figure}
  \resizebox{\hsize}{!}{\includegraphics[angle=0]{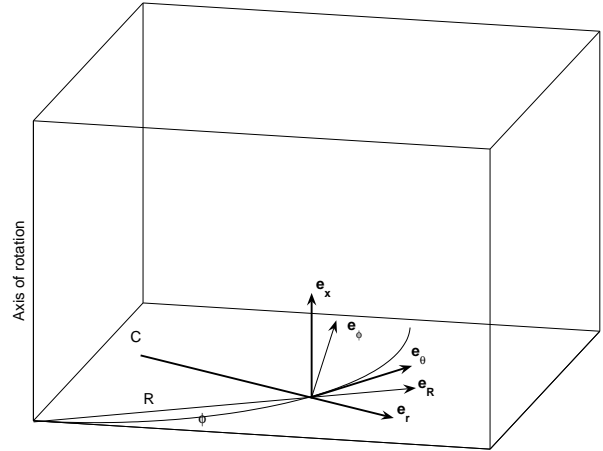}}
  \caption{Two orthonormal bases are
considered: i) cylindrical components of unit vectors, (${\bf
e}_\Phi, {\bf e}_R$, ${\bf e}_x$); ii) unit vectors of the system
rigidly fixed on each point of the curve, (${\bf e}_r, {\bf
e}_{\theta}$, ${\bf e}_x$), respectively. $C$ is the center of the
curvature. Hereafter, this set of coordinates is referred to as the
field line coordinates.}\label{fig}
\end{figure}

\section{Main consideration} \label{sec:consid}
%
%
%
We begin our investigation by considering the two-component plasma
consisting of the relativistic electrons with the Lorentz factor
$\gamma_{e}\sim 10^{5-8}$ (see \cite{osm7,ra8}) and the bulk
component (protons) with $\gamma_{b}\sim 10$. Since we are
interested in the twisting process, we suppose that initially the
field lines are almost rectilinear to study how this configuration
changes in time.

\begin{figure}
  \resizebox{\hsize}{!}{\includegraphics[angle=0]{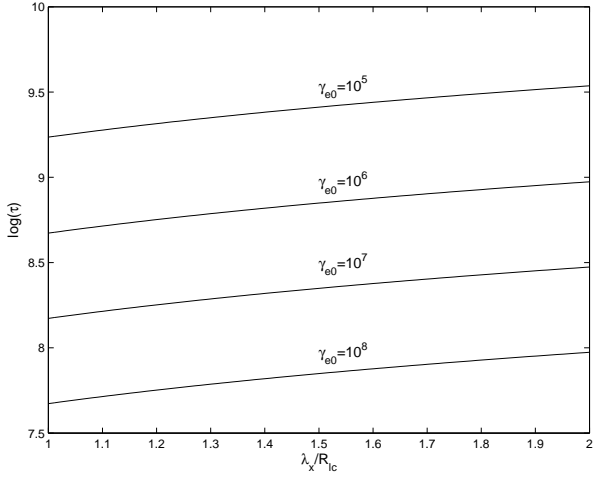}}
  \caption{The dependence of logarithm of the instability timescale
  on the normalized wave length. The set of parameters is
  $\gamma_{e0} = \{10^5;10^6;10^7;10^8\}$, $R_B\approx R_{lc}$, $n_{e0} =
0.001cm^{-3}$, $\lambda_{\phi} = 100R_{lc}$, $L/L_E = 0.01$, $L_E =
10^{46}erg/s$ is the Eddington luminosity for the given AGN
mass.}\label{lambda}
\end{figure}

We express the equation of motion in the cylindrical coordinates
(see Fig. \ref{fig}) and we start by considering the Euler equation:

\begin{equation}
\label{eul} \frac{\partial{\bf p_{\alpha}}}{\partial t}+({\bf
v_{\alpha}\nabla)p_{\alpha}}=
-c^2\gamma_{\alpha}\xi{\bf\nabla}\xi+\frac{q_{\alpha}}{m_{\alpha}}\left({\bf
E}+ \frac{1}{c}{\bf v_{\alpha}\times B}\right),
\end{equation}
the continuity equation:

\begin{equation}
\label{cont} \frac{\partial n_{\alpha}}{\partial t}+{\bf
\nabla}(n_{\alpha}{\bf v_{\alpha}})=0,
\end{equation}
and the induction equation:
\begin{equation}
\label{ind} {\bf \nabla\times B} = \frac{1}{c}\frac{\partial {\bf
E}}{\partial t}+\frac{4\pi}{c}\sum_{\alpha=e,b}{\bf J_{\alpha}},
\end{equation}
where $\alpha=\{e,b\}$, $ \label{xi} \xi\equiv
\sqrt{1-\Omega^2R^2/c^2}, $, ${\bf p_{\alpha}}$ is the momentum,
${\bf v_{\alpha}}$ - the velocity and ${\gamma_{\alpha}}$ is the
Lorentz factor of the relativistic particles.

We linearize the system of equations Eqs. (\ref{eul}-\ref{ind}),
perturbing all physical quantities around the leading state
$\Psi\approx \Psi^0 + \Psi^1$, where $\Psi = \{n,{\bf v},{\bf
p},{\bf E},{\bf B}\}$ and $\Psi^1(t,{\bf r})\propto\Psi^1(t)
\exp\left[i\left({\bf kr} \right)\right]$.

Then, from Eqs. (\ref{eul}-\ref{ind}) we derive the linearized set
of equations governing the CDI:

\begin{figure}
  \resizebox{\hsize}{!}{\includegraphics[angle=0]{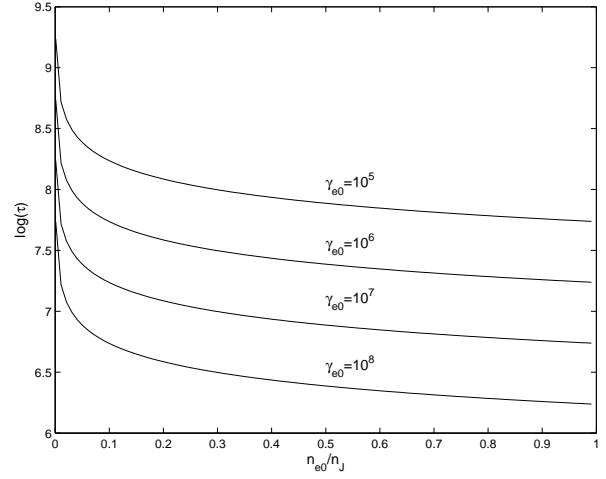}}
  \caption{The dependence of logarithm of the instability timescale
  on the density normalized by the medium density. The set of parameters is
  $\gamma_{e0} = \{10^5;10^6;10^7;10^8\}$, $R_B\approx R_{lc}$, $\lambda_{e0} =
R_{lc}$, $\lambda_{\phi} = 100R_{lc}$ and $L/L_E = 0.01$. Here,
$n_{e0}$ is normalized by the intergalactic medium density,
$n_m\approx 1cm^{-3}$}\label{density}
\end{figure}

\begin{equation}
\label{eulp} \frac{\partial p^1_{{\alpha}x}}{\partial
t}-i(k_xu_{\alpha}+k_{\phi}v_{_\parallel})p^1_{{\alpha}x}=
\frac{q_{\alpha}}{m_{\alpha}}v_{_\parallel}B^1_{r},
\end{equation}
\begin{equation}
\label{contp} \frac{\partial n^1_{\alpha}}{\partial
t}-i(k_xu_{\alpha}+k_{\phi}v_{_\parallel})n^1_{\alpha}=
ik_xn_{\alpha}^0v^1_{\alpha x},
\end{equation}
\begin{equation}
\label{indp} -ik_{\phi}cB^1_{r} = 4\pi
\sum_{\alpha=e,b}q_{\alpha}(n_{\alpha}^0v^1_{\alpha
x}+n_{\alpha}^1u_{\alpha}).
\end{equation}
By $u_{\alpha}= \gamma_{\alpha_0} v_{_\parallel}^2/(\omega_B R_B)$,
we denote the curvature drift velocity along the $x$ axis, where
$\omega_{\alpha B} = q_{\alpha}B_0/m_{\alpha}c$, $R_B$ is the
curvature radius of magnetic field lines, $v_{_\parallel} \approx
c\cos(\Omega t)$, $B_0 = \sqrt{2L/(R_{lc}c^2)}$,  $L$ is the
luminosity of the AGN and $R_{lc}=c/\Omega$ - the light cylinder
radius. In deriving Eqs. (\ref{eulp}-\ref{indp}), the wave
propagating almost perpendicular to the equatorial plane, was
considered and the expression $v^1_r\approx cE^1_x/B_{0}$ was taken
into account. For simplicity, the set of equations are given in
terms of the coordinates of the field line (see Fig. \ref{fig}).

We express $v^1_{\alpha x}$ and $n^1_{\alpha}$ in the following way:

\begin{equation}
\label{anzp} v^1_{\alpha x}\equiv V_{\alpha x}e^{i{\bf
kA_{\alpha}(t)}},
\end{equation}
\begin{equation}
\label{anzn} n^1_{\alpha}\equiv N_{\alpha}e^{i{\bf kA_{\alpha}}(t)},
\end{equation}
where
\begin{equation}
\label{Ax} A_{\alpha x}(t) = \frac{u_{\alpha}t}{2} +
\frac{u_{\alpha}}{4\Omega}\sin(2\Omega t),
\end{equation}
\begin{equation}
\label{Af} A_{\alpha\phi}(t) = \frac{c}{\Omega}\sin(\Omega t).
\end{equation}
Then, by substituting Eqs. (\ref{anzp}) and (\ref{anzn}) into Eqs.
(\ref{eulp}-\ref{indp}), it becomes straightforward to solve the
system for the toroidal component and find a corresponding increment
of the instability:

\begin{equation}
\label{increm} \Gamma \approx
\left(-\frac{3}{2}\frac{\omega^2_{e}}{\gamma_{e_0}}\frac{k_xu_{e}}{k_{\theta}c}\right)
^{1/2}\left|J_0\left(\frac{k_xu_{e}}
{4\Omega}\right)J_{0}\left(\frac{k_{\theta}c}{\Omega}\right)\right|.
\end{equation}

\begin{figure}
  \resizebox{\hsize}{!}{\includegraphics[angle=0]{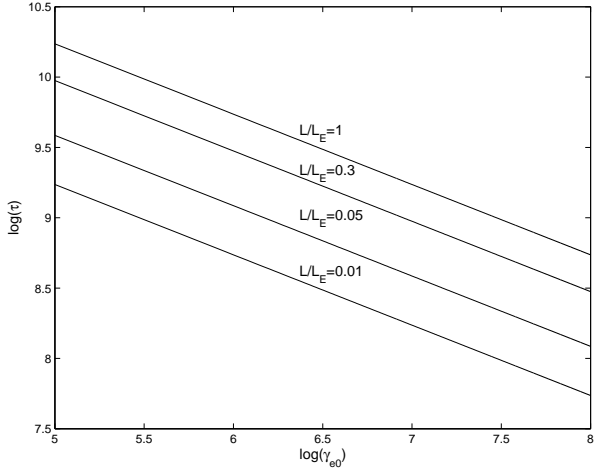}}
  \caption{The dependence of logarithm of the instability time scale
  on $log(\gamma_{e0})$. The set of parameters is
  $R_B\approx R_{lc}$, $n_{e0}/n_m = 0.001$, $\lambda_{e0} =
R_{lc}$, $\lambda_{\phi} = 100R_{lc}$ and $L/L_E =
\{0.01;0.05;0.3;1\}$.}\label{gamma}
\end{figure}

\section{Results} \label{sec:results}
%
%
%
We investigate the CDI growth rate in terms of the wavelength, the
density of relativistic electrons, their Lorentz factors and the AGN
bolometric luminosity.

In studying the behaviour of the instability as a function of the
wavelength, we examine the typical AGN parameters:
$M_{BH}=10^{8}\times M_{\odot}$, $\Omega = 5\times 10^{-5}s^{-1}$
and $L = 10^{44}erg/s$, where $M_{BH}$ is the AGN mass, $M_{\odot}$
is the solar mass and $L$ is the bolometric luminosity of the AGN.

We consider Eq. (\ref{increm}) and plot the logarithm of the
instability timescale $\tau\equiv 1/\Gamma$ versus the wavelength.
The present consideration is based on the centrifugal acceleration.
As shown in \cite{osm7}, due to the CF, the relativistic particles
can reach very high Lorentz factors. For this purpose, it is
reasonable to investigate the efficiency of the instability in terms
of the wavelength but for different values of Lorentz factors. Fig.
\ref{lambda} shows the mentioned behaviour for different parameters.
Different curves correspond to different values of Lorentz factors.
For the given range of $\lambda_x$ and different values of
$\gamma_{e0}$, the CDI time scale varies from $\sim 10^7s$
($\lambda_{x}/R_{lc} = 1$, $\gamma_{e0} = 10^8$) to $\sim 10^9$
($\lambda_{x}/R_{lc} = 2$, $\gamma_{e0} = 10^5$).

In Fig. \ref{density} the plots of $log(\tau)$ versus the AGN wind
density illustrate that the timescale is a continuously decreasing
function of $n_{e0}/n_m$. As we see from the figure, $\tau$ varies
from $\sim 10^9s$ ($n_{e0}/n_m = 0.001$, $\gamma_{e0} = 10^5$) to
$\sim 10^6s$ ($n_{e0}/n_m = 1$, $\gamma_{e0} = 10^8$).

In Fig. \ref{gamma}, the plot of $log(\tau)$ versus
$log(\gamma_{e0})$ is shown for different luminosities. As we see,
the instability timescale varies from $\sim 10^{10}s$ ($\gamma_{e0}
= 10^5$, $L/L_E=1$) to $\sim 10^8s$ ($\gamma_{e0} = 10^7$,
$L/L_E=0.01$). On the other hand, the plots for different
luminosities illustrate another property of $\tau$: by increasing
the luminosity of the AGN, the corresponding instability becomes
less efficient.

\begin{figure}
  \resizebox{\hsize}{!}{\includegraphics[angle=0]{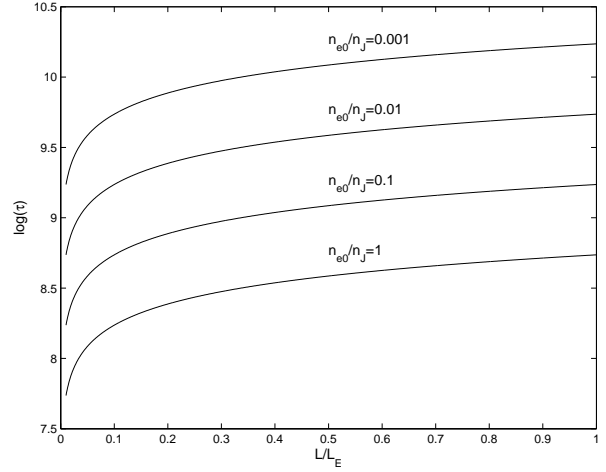}}
  \caption{The dependence of logarithm of the instability time scale
  on $log(\gamma_{e0})$. The set of parameters is $R_B\approx
  R_{lc}$, $n_{e0}/n_m = \{0.001;0.01;0.1;1\}$, $\lambda_{e0} =
R_{lc}$, $\lambda_{\phi} = 100R_{lc}$ and $L/L_E =
0.01$.}\label{lumin}
\end{figure}

To observe this particular feature more clearly, we consider how in
Fig. \ref{lumin} the dependence of $log(\tau)$ on $L/L_E$ is clearly
evident for different values of densities. From the plots, it is
seen that by increasing the luminosity, the timescale continuously
increases. For the afore mentioned area of quantities, the timescale
varies from $\sim 10^7s$ ($L/L_E = 0.01$, $n_{e0}/n_J = 1$) to $\sim
10^{10}s$ ($L/L_E = 1$, $n_{e0}/n_J = 0.001$).

We observe from the present investigation that the instability
timescale varies in the following range: $\tau\in\{10^6;10^{10}\}s$.
To specify how efficient the CDI is, it is pertinent to examine an
accretion process, estimate its corresponding evolution timescale,
and compare this value with that of the CDI.

Considering the problem of fuelling AGNs \cite{king} it was showed
that the accretion timescale can be estimated by the following form
$t_{evol}  =3.53\times
10^{13}\left(10L/L_E\right)^{-22/27}\left(10^{-8}M_{BH}/M_{\odot}\right)^{-4/27}s$.
As is clear from this formula, the accretion evolution timescale
depends on two major AGN parameters, the luminosity ($L$) and the
AGN mass ($M_{BH}$). Therefore, it is reasonable to investigate
$t_{evol}$ versus $L$ and $M_{BH}$. Let us examine the following
ranges of variables: $M_{9}\equiv M_{BH}/(10^9\times
M_{\odot})\in\{0.01;1\}$ and $L/L_E\in\{0.01;1\}$. Then it is easy
to show that the minimum value of the evolution timescale is of the
order of $\sim 10^{12}s$, which corresponds to $M_9 = 1$ and $L/L_E
= 1$, whereas the maximum value, approximately $\sim 10^{15}s$,
corresponds to the following pair of variables $M_9 = 0.001$ and
$L/L_E = 0.01$.

As has been found, $\tau$ varies in the range $\sim
(10^{6}-10^{10})s$, whereas the sensible area of $t_{evol}$ is $\sim
(10^{12}-10^{15})s$. Therefore, the instability timescale is less
than the evolution timescale of the accretion by many orders of
magnitude, which implies that the linear stage of the CDI is
extremely efficient.

The twisting process of magnetic field lines requires a certain
amount of energy and it is natural to study the energy budget of
this process. For this reason we have to introduce the maximum of
the possible luminosity $L_{max} = \dot{M}c^2$ and compare this with
the "luminosity" corresponding to the reconstruction of the magnetic
field configuration $L_m\equiv \Delta E_m/\Delta t\approx\Delta
E_m/\tau$, where $\Delta E_m$ is the variation in the magnetic
energy due to the curvature drift instability.

We consider a AGN of the luminosity, $L = 10^{45}erg/s$, then, the
accretion provides the following maximum value: $L_{max} =L/0.1=
10^{46}erg/s$. On the other hand, if the process of sweepback is
realistic, the magnetic "luminosity" cannot exceed $L_{max}$. The
magnetic "luminosity" can be expressed by following $L_{m} =
B_r^2R_{lc}^3\kappa/(4\pi\tau)$, with $B_r = B_r^0exp(t/\tau)$,
where $B_r^0$ is the initial perturbation of the toroidal component
of magnetic field and $\kappa\equiv \Delta R/R_{lc}<<1$ represents
the non-dimensional thickness of a thin spatial layer close to the
LCS.

We introduce the initial non-dimensional perturbation, $\delta$,
defined to be $\delta\equiv B_r^0/B_0$, where by $B_0$ we denote the
induction of the magnetic field in the leading state. By considering
the following set of parameters $\gamma_{e0} =
\{10^5;10^6;10^7;10^8\}$, $R_B\approx R_{lc}$, $n_{e0} =
0.001cm^{-3}$, $\lambda_{\phi} = 100R_{lc}$, $\lambda_{x} = R_{lc}$
and $L = 10^{45}erg/s$, we investigate the behaviour of
$L_{m}/L_{max}$ versus the initial perturbation for the
characteristic timescale ($t\approx\tau$). One can see that,
$L_{m}/L_{max}$ varies from $\sim 0$ ($\delta = 0$) to $\sim
2.3\times 10^{-9}$ ($\delta = 0.1$, $\gamma_{e0} = 10^8$).
Therefore, $L_{max}>>L_{m}$ which means that only a tiny fraction of
the total energy goes to the sweepback, making this process
feasible.

\section{Summary} \label{sec:summary}
%
%
%

We summarize the principal steps and conclusions of our study to be:

\begin{enumerate}

      \item Considering the relativistic two-component plasma for
      AGN winds, the centrifugally driven
      curvature drift instability has been studied.

      \item Taking into account a quasi single approach for the particle dynamics,
      we linearized the Euler continuity and induction equations. The dispersion
      relation characterizing the parametric instability of the toroidal component of the magnetic
      field has been derived.

      \item By considering the proper frequency of the curvature drift modes,
      the corresponding expression of the instability increment has been obtained for
      the light cylinder region.

      \item The efficiency of the CDI has been investigated
      by adopting four physical parameters, namely: the wavelength,
      flow density and Lorentz factors of electrons, and the luminosity
      of AGNs.

      \item By considering the evolution process of accretion,
      the corresponding timescale has been estimated for a
      physically reasonable area in the parametric space
      $L-M_{BH}$. It was shown that the instability timescale was
      lower by many orders of magnitude than the evolution
      timescale, indicating extremely high efficiency of the CDI.

      \item Examining the instability from the point of view of
      the energy budget, we have seen that the sweepback of the
      magnetic field lines requires only a small fraction of the
      total energy, which means that the CDI is a realistic
      process.

\end{enumerate}

\begin{theacknowledgments}
I thank professor G. Machabeli for valuable discussions. The
research was supported by the Georgian National Science Foundation
grant GNSF/ST06/4-096.
\end{theacknowledgments}



\end{document}